\documentclass[runningheads]{llncs}

\usepackage{graphicx}
\usepackage{amsmath}
\usepackage[caption=false]{subfig}
\usepackage{cite}

\newcommand{\yb}{\mathbf{y}}
\newcommand{\zb}{\mathbf{z}}

\begin{document}

\title{Simulation of virtual cohorts increases predictive accuracy of cognitive decline in MCI subjects}

\titlerunning{Simulation of virtual cohort increases predictive accuracy}

\author{Igor Koval \inst{1} \and St\'ephanie Allassonni\`ere \inst{2} \and Stanley Durrleman \inst{1}, for the Alzheimer's Disease Neuroimaging Initiative }
\authorrunning{Igor Koval et al.}
\institute{Inria Paris-Rocquencourt, Inserm U1127, CNRS UMR 7225, Sorbonne Universit\'{e}s, UPMC Univ Paris 06 UMRS 1127, Institut du Cerveau et de la Moelle \'{e}pini\`{e}re, ICM, F-75013, Paris, France \and INSERM UMRS 1138, Centre de Recherche des Cordeliers, Université Paris Descartes, Paris, France}
\maketitle              % typeset the header of the contribution
\begin{abstract}
The ability to predict the progression of biomarkers, notably in NDD, is limited by the size of the longitudinal data sets, in terms of number of patients, number of visits per patients and total follow-up time. To this end, we introduce a data augmentation technique that is able to reproduce the variability seen in a longitudinal training data set and simulate continuous biomarkers trajectories for any number of virtual patients. Thanks to this simulation framework, we propose to transform the training set into a simulated data set with more patients, more time-points per patient and longer follow-up duration. We illustrate this approach on the prediction of the MMSE of MCI subjects of the ADNI data set. We show that it allows to reach predictions with errors comparable to the noise in the data, estimated in test/retest studies, achieving a improvement of 37\% of the mean absolute error compared to the same non-augmented model. 

\end{abstract}
\section{Introduction}

Predicting the future progression of patients with neurodegenerative diseases (NDD) is a key challenge to treat patients at an earlier stage than today or to better evaluate drug efficacy in clinical trials. Longitudinal data sets, consisting of repeated observations of the same patients over time, play a central role to describe and predict disease progression. Machine Learning techniques, trained on sequence data (multiple observations per patient), have seize this challenge. However, the databases often lack patients with sufficient follow-up visits, a future visit to predict and a sufficiently large delay in between, leading to poor generalization on test data. More importantly, the different experimental settings - time to prediction or patients/visits presenting the considered feature(s) - are difficult to compare as they involve a subset of the initial cohort that has specific characteristics in terms of number of patients, number of follow-up visits and duration. It is often impossible to balance the training and test set for all these characteristics.  This problem prevents from a reliable comparison of the algorithms, some being better due to the size or characteristics of the training set, rather than the intrinsic performance of the algorithm or choice of the features.%, which is critical when studying the importance of new biomarkers. 

To increase the size of real data sets, data augmentation techniques have been developed : virtual data are drawn with the intention to reproduce the characteristics of the data in the initial cohort. Most of the literature focuses on techniques for independent and identically distributed observations such as image classification \cite{perez2017effectiveness}, text categorization \cite{lu2006enhancing} or speech recognition \cite{cui2015data}. Due to the unrealistic hypothesis for sequence data, some techniques have been proposed for uni-dimensional time-series. They rely on a continuous transformation of the time domain by warping, slicing or sliding the time window \cite{le2016data}. Such techniques do not apply to NDD as the temporal pattern is key in the disease progression. Recently, Generative Adversarial Networks (GAN) have received interest due to the characteristic of the generative part of the model: it can sample virtual realistic data. It is however non-trivial to generate sequence data as there is no straightforward way to propagate the gradient updates from the discriminator to the generator \cite{yu2017seqgan}. Furthermore, these models rely on large training data sets, which are typically inaccessible in the targeted medical applications.

In this paper, we propose to use a simulation framework to perform data augmentation for sequence data in the presence of small training samples, and to evaluate to which extend it increases the performance of a predictive algorithm. The model introduced in \cite{schiratti2017bayesian} recombines short-term individual observations at different disease stages to estimate a long-term scenario of disease progression. The model estimates also parameters that change the pattern of progression, the pace of progression and the age at onset, so that it can reconstruct a continuous trajectory by fitting individual data, or simulate entirely synthetic trajectories by sampling the empirical distribution of the parameters. This model, once trained on a small and unbalanced training data set, may be used therefore to re-sample the trajectories of the training subjects by adding new visits and covering larger time span, and even increase the number of subjects by adding data sampled from simulated trajectories. This augmented virtual cohort may then be used in lieu of the training samples to train a predictive algorithm. We propose to evaluate the performance of the algorithm trained with this augmented and virtualized data, as compared to the original training set, for the prediction of the cognitive decline in subjects with mild cognitive impairments.

%To address the problem of data augmentation for sequential data in the presence of small data sets, we introduce a technique that is able to accurately simulate individual trajectories with an arbitrary number of visits. It is based on a model introduced in \cite{schiratti2017bayesian} that recombine short-term individual observations of different disease stages to estimate the group-average long-term progression of the phenomenon at hand. At the same time, the model estimates the spatio-temporal variability of disease progression in the form of a distribution of individual parameters. This distribution can be used to draw new samples, that characterize new individual progressions, with little constraints on the number of observations per subject or the time between each visit. The simulated patients can then be used as training patients to enhance predictive models, with potentially more follow-up visits to build the prediction on. 

%We evaluate this method for the long-term prediction of cognitive decline in subjects with Mild Cognitively Impairements (MCI) from the Alzheimer's Disease Neuroimaging Initiative (ADNI) database. We demonstrate that the technique developped in \cite{schiratti2017bayesian} allows, first, to achieve a significant improvement for the prediction of the Mini-Mental State Examination (MMSE) over 4 years following the last observations of the patient, and secondly, to compare the predictive power of different sets of variables.

\begin{figure}[bt!]
\includegraphics[width=\textwidth]{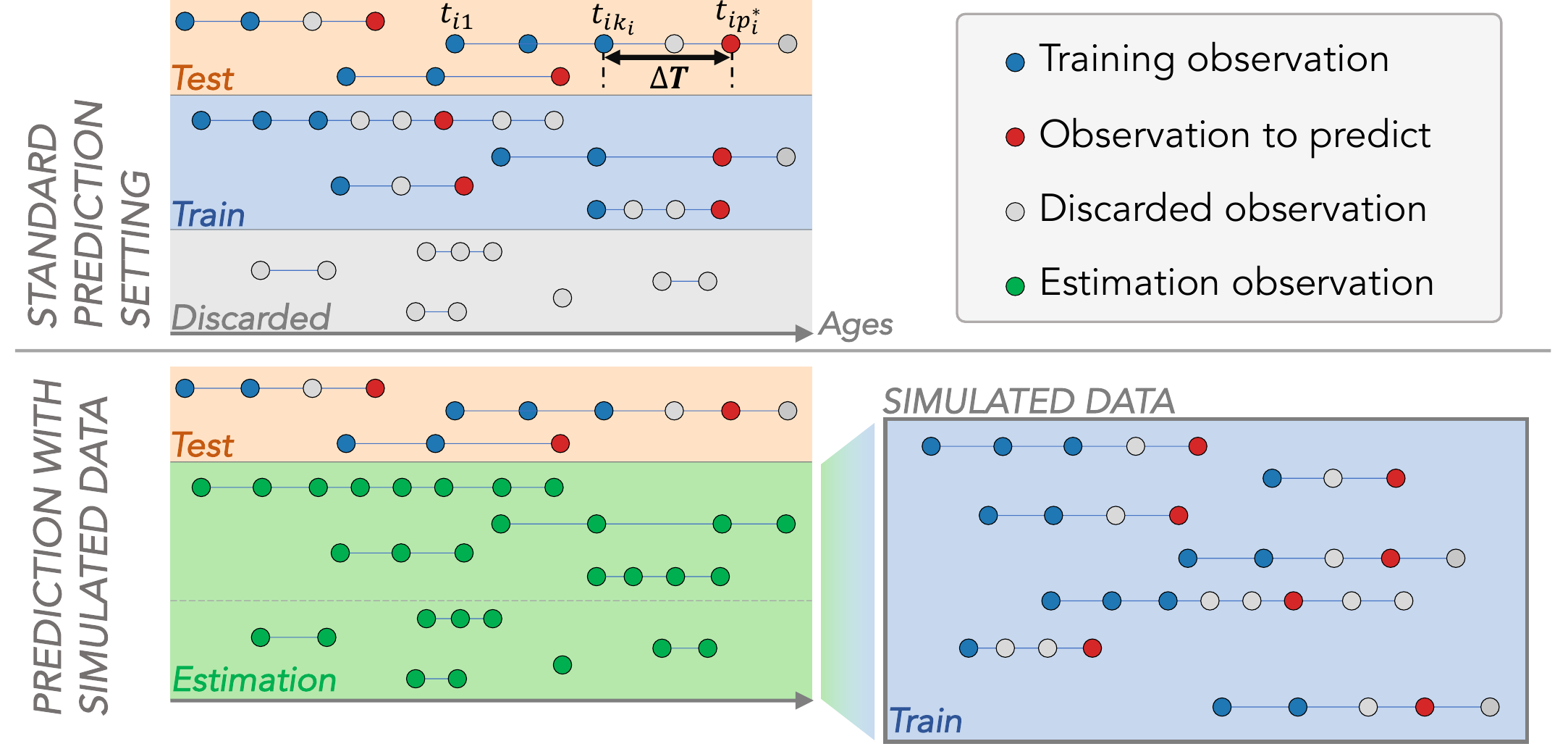}
\caption{The top row describes the standard prediction setting where the data set is split in a train and test set. A fixed time-delay $\Delta T$ is set between the input visits (blue dots) and the target prediction (red dot). It leads to discarding visits in between (grey dots) or entire subjects that do not present sufficient follow up visits (discarded set). The bottom row corresponds to the procedure with simulated data : the training set is composed of virtual patients that are simulated thanks to the estimation set.} \label{fig:setting}
\end{figure}

\section{Sequence-based prediction}

In the following, we consider a longitudinal data set $\yb = (\yb_{ij}, t_{ij})_{\substack{1 \leq i \leq n \\ 1 \leq j \leq n_i}}$ where the $i-$th subject has been observed $n_i$ times at ages $ t_{1i} < \dots < t_{i n_i}$ with $\yb_{ij} \in R^d$ a set of biomarkers. Each observation corresponds to a snapshot of the individual spatiotemporal trajectory. We aim to predict the value of one biomarker in $\Delta T$ years after a given visit, knowing the values of the biomarkers at the previous visits.

\subsection{Standard prediction setting}

In a standard setting, one needs to discard patients that do not have sufficient follow-up visits to cover the required temporal span, i.e. $t_{i n_{i}} - t_{i1} < \Delta T$, as shown on the top row of Figure \ref{fig:setting}. For the remaining patients, the input biomarkers $(\yb_{ij})_{1 \leq j \leq k_i}$ at some early visits  $(t_{ij})_{1 \leq j \leq k_i}$ (blue dots) are used to predict the biomarker $\yb_{i p_{i}^*}$ at age $t_{i p_{i}^*}$ (red dot) such that $t_{i p_{i}^*} = t_{i k_i} + \Delta T$. This task may be achieved by any machine learning algorithm, for instance a neural network. In this setting, possible intermediate visits $t_{ij}$ such that $t_{ik_i} < t_{ij} < t_{i p_{i}^* }$ are discarded (grey dots). If multiple splits of input/output visits $((\yb_{ij}, t_{ij})_{1 \leq j \leq k_i}, (\yb_{i p_{i}^*}, t_{i p_{i}^*}))$ are possible, one is selected at random. Once the input and target visits have been selected for each patient, they are split into the train and test set.

The longer $\Delta T$, the fewer patients remain in the train and test set and the fewer visits per remaining patients there are. Therefore, when  $\Delta T$ is varied, the size and composition of the train and test set may vary dramatically, and thus the performance of the predictive algorithm. This problem is even more critical if several biomarkers that are not observed at every visit are used. %it is not straightforward to compare the models as they will rely on different training/testing settings. Furthermore, if not enough patients remain, it becomes difficult to discriminate between a model that benefits from the full potential of the variable(s) at hand and a model that simply lacks patients to get better results. 

%It should be mentioned that contrary to what has just been written for clarity purposes, there is no need to use all the coordinates of $(\yb_{ij})_{1 \leq j \leq k_i}$ neither to predict all the coordinates of $\yb_{i p_{i}^*}$ : it is possible to select a subset of the $d$ features ($\yb_{ij} \in R^d$) to predict another one.

\subsection{Sequence data simulation}

We consider a generative mixed-effect model such that the individual observations at time-point $t_{ij}$ is $\yb_{ij} = f(\theta, \zb_i, t_{ij}) + \epsilon_{ij} \quad \text{,}  \, \,  \epsilon_{ij} \overset{\text{i.i.d.}}{\sim} \mathcal{N}(\mathbf{0}, \sigma^2 \text{Id}_{d})$ where $\theta$ corresponds to the fixed-effects and $\zb_i$ the random effects associated to the subject $i$. Assuming that it is possible to estimate $\theta$ and  $(\zb_i)_{1 \leq i \leq n}$ given a training data set, we can draw a new sample $\zb_{i'}$ from the empirical distribution of the  $(\zb_i)_{1 \leq i \leq n}$. It corresponds to a new individual for which is it possible to simulate new observations at arbitrary time-point $(t_{i'j})_{1 \leq j \leq n_{i'}}$.

An example of such model is introduced in \cite{schiratti2017bayesian}, where the authors consider that the fixed-effects $\theta$ define the group-average disease progression and its variability in the population. This fixed-effects, of small dimension relatively to the feature space, can be estimated, thanks to the MCMC-SAEM \cite{delyon1999, allassonniere2015}, with short-term observations for a relatively small number of subjects. As for the $i$-th individual trajectory, the authors consider that it derives from the group-average trajectory according to the random-effects $\zb_i = (\alpha_i, \tau_i, (s_{ij})_{1 \leq j \leq N_s})$ where $\alpha_i$ corresponds to the pace of progression and $\tau_i$ relates about the time delay between the group-average and the individual scenario. On top of these temporal parameters that impact the observation coordinates similarly, the space-shifts $(s_{ij})_{1 \leq j \leq N_s}$ characterize the inter-coordinates variations ($N_s \leq N$). These random-effects are learnt by optimizing the individual complete likelihood $p((\yb_{ij})_{1 \leq j \leq n_i}, \zb_i ; \theta) = p((\yb_{ij})_{1 \leq j \leq n_i} | \zb_i ; \theta) p(\zb_i ; \theta)$ thanks to the L-BFGS-B method \cite{zhu1997algorithm}. %The first term corresponds to the data attachment while the second is a regularization term that comes from the prior over the hidden variables as the model is described in a Bayesian setting.
 
 %The procedure results in an empirical distribution of the individual parameters $(\zb_i)_{1 \leq i \leq n}$. 
 It is possible to draw a set of individual parameters $\zb_{i'}$ by, first, simulating the temporal parameters $(\alpha_{i'}, \tau_{i'})$ with a kernel density estimation on the empirical distribution $(\alpha_{i}, \tau_{i})_{1 \leq i \leq n}$. Then, considering the multivariate Gaussian distribution $\mathcal{N}(\mathbf{\mu}, \mathbf{\Sigma})$ estimated on the whole learnt random effects $(\zb_i)_{1 \leq i \leq n}$, it is possible to draw $((s_{i'j})_{1 \leq j \leq N_s} | \alpha_{i'}, \tau_{i'}) \sim \mathcal{N}(\mathbf{\tilde{\mu}}, \mathbf{\tilde{\Sigma}})$ where $\mathbf{\tilde{\mu}}$ and $\mathbf{\tilde{\Sigma}}$ are functions of $\mathbf{\mu}$ and $\mathbf{\Sigma}$ \cite{petersen2008matrix}.

 %As describe in eq. \ref{eq:generic_model}, the subject $i'$ is fully described by $\theta$ and $\zb_{i'}$. The ages at which (s)he is observed $(t_{i'k})_{1 \leq k \leq n_{i'}}$ are only constrained by the fact that ages that have not been seen in the real cohort (modulus the time reparametrization implied by the model) are not likely to be reproduced correctly. Said differently, it is hard to generate time spans that are not present in the initial cohort e.g. before or after the phenomenon described in the original data set. Apart from that, the model allows to select an arbitrary number of follow-ups with an arbitrary time-step between them.

\subsection{Prediction setting with simulated patients}

From the whole data set, we first select an estimation subset that includes the subjects that were discarded in the standard procedure along with some subjects that have more follow-up duration, as shown on the bottom row of Figure \ref{fig:setting}. The remaining subjects form the test set, with same split constraints on the input and target visits as in the standard prediction setting.

Once $\theta$ and $(\zb_i)_{1 \leq i \leq n}$ are learnt on the estimation set, we can simulate an arbitrary number of patients and visits per individual that will be used as the training set. To have similar characteristics in the training and test set, the time between two simulated visits is kept similar to the one of the real patient (e.g. one year), and, the input and target visits are split as before. It is important to mention that the estimation procedure now uses the visits $t_{ij}$ such that $t_{ik_i} < t_{ij} < t_{ij^*}$ that were previously discarded (grey dots). 

In the following, to assess the quality of the simulated data only, we prevent ourselves from using patients from the estimation set that are eligible in term of number of visits to be used in the training set (upper part of the estimation set of Figure \ref{fig:setting}) : that the training procedure relies on the simulated patients only. In other cases, it is indeed possible - and recommended - to add them to the training set.

\begin{figure}[h]
  \begin{center}
    \subfloat[Normalized ADAS 11]{
      \includegraphics[width=0.32\textwidth]{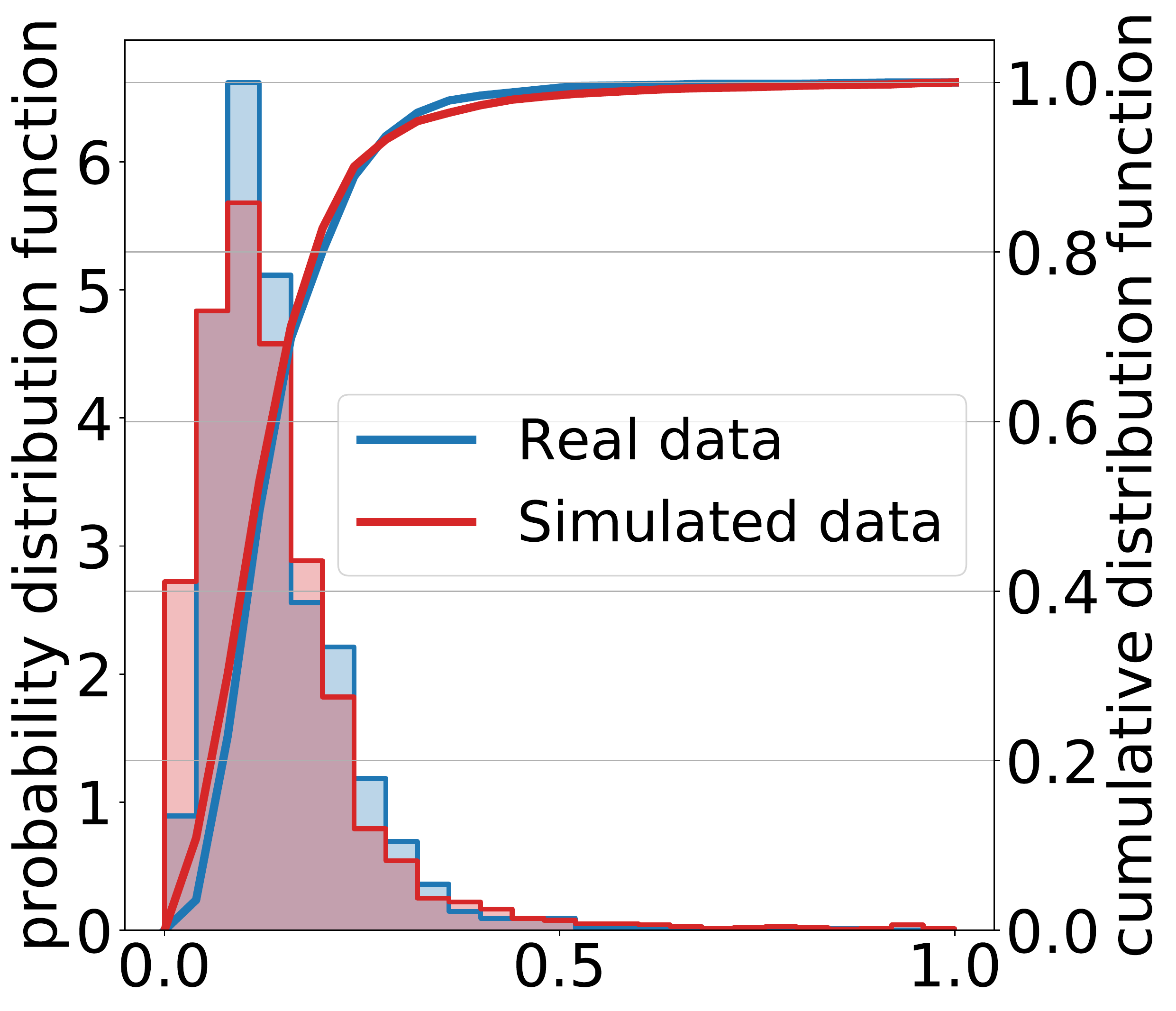}
      \label{fig:sub-ADAS}
                         }
    \subfloat[Normalized MMSE]{
      \includegraphics[width=0.32\textwidth]{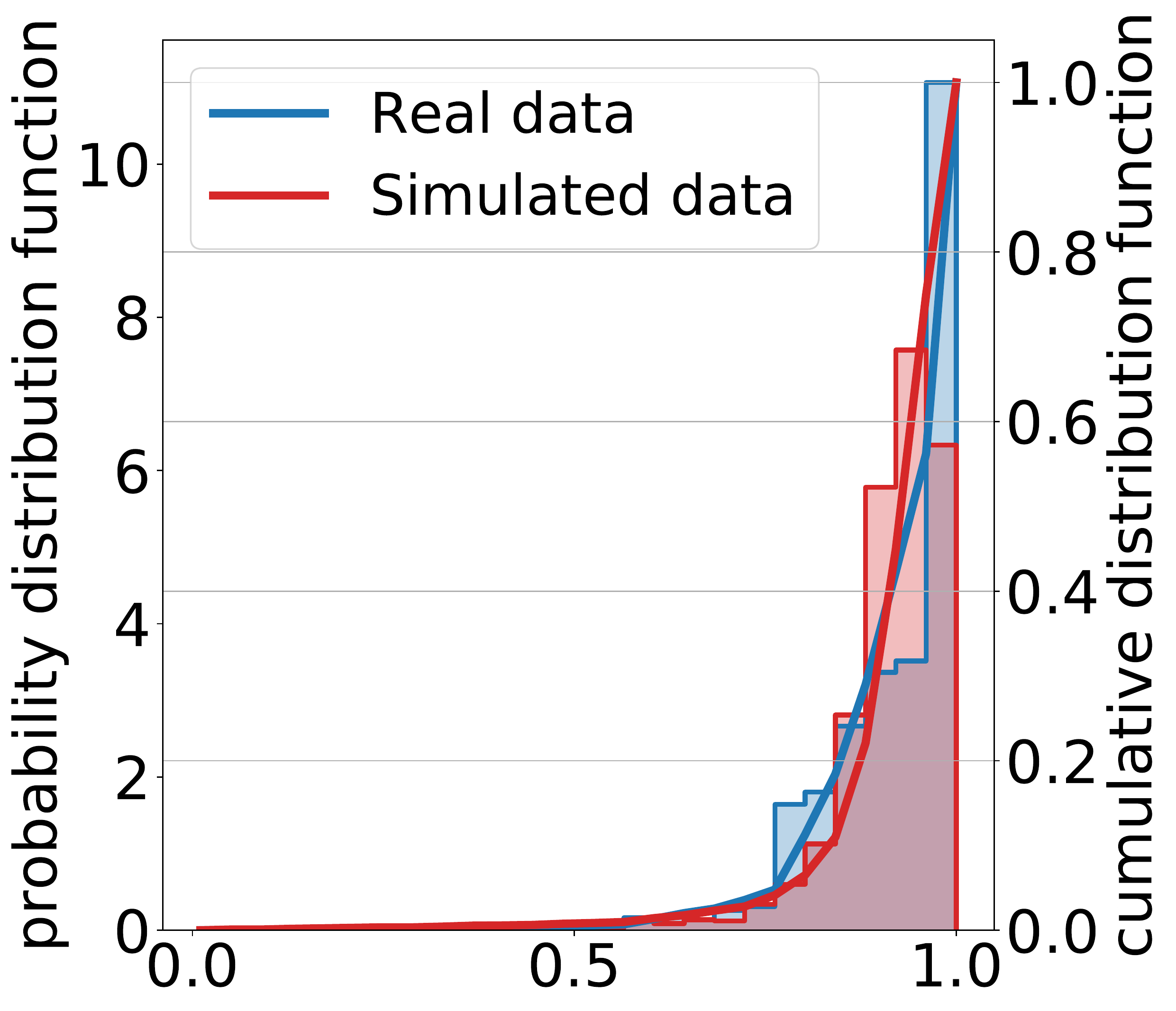}
      \label{fig:sub-MMSE}
                         }
    \subfloat[Normalized MOCA]{
      \includegraphics[width=0.32\textwidth]{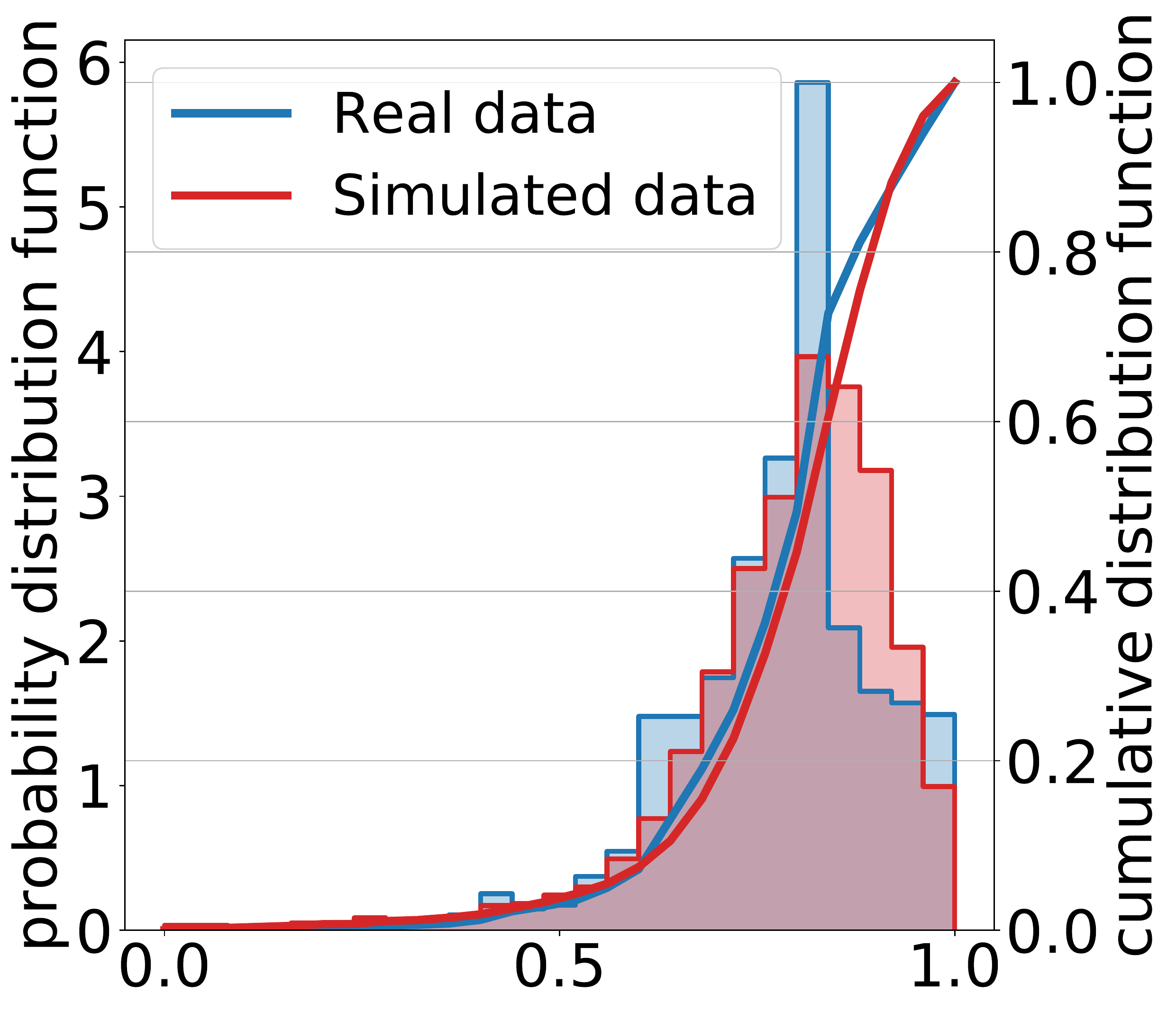}
      \label{fig:sub-MOCA}
                         }
    \caption{Empirical histograms and cumulative distributions of original cohort (blue) and of the simulated data (red) based on a model that estimate the evolution of the normalized ADAS Cog with 11 and 13 items and the normalized MOCA.}
    \label{fig:data-simulation}
  \end{center}
\end{figure}

\section{Experimental results}

\subsection{Data and experimental setting}

The experiments focus on the prediction of the mini-mental state examination (MMSE) for subjects with mild cognitive impairment (MCI) from the ADNI database. Therefore, our cohort includes early and late MCI, but also MCI who converted to Alzheimer's Disease and stable MCI. We considered predictions at 1, 2, 3 and 4 years, based on a set of features among the MMSE, Alzheimer's disease assessment scale - cognitive subscale (ADAS-Cog) with 11 or 13 items, clinical dementia rating sum of boxes (CDRSB), the Montreal cognitive assessment (MOCA) and the functional assessment questionnaire (FAQ). One estimation set was defined per subset of features used in the predictive model, e.g. an estimation set that simulate patients with MMSE, ADAS-11 and ADAS-13 was defined for the predictions that are based on these features.

To assess the performance of the simulation framework, we create a virtual cohort with the same characteristics (number of patients and time-points) as the estimation set. The empirical data distribution for the real and virtual cohort are nearly identical as shown on Figure~\ref{fig:data-simulation} for three different features.

We choose a long short-term memory (LSTM) neural network, with 10 hidden dimensions, stacked with a linear layer, as our algorithm to predict future feature values. The Mean Squared Error (L2-norm) loss is optimized thanks to the ADAM optimizer (learning rate of $10^{-3}$ and weighted decay of $10^{-5}$). To prevent the model from overfitting, a subset of the real patients, namely the validation set, is used to apply the early stopping criterion procedure : it stops the training if no loss improvement is detected from a given number of epoch on the validation set. The code is available at [shown after paper acceptance]. 

\begin{figure}[h]
  \begin{center}
    \subfloat[Standard prediction]{
      \includegraphics[width=0.5\textwidth]{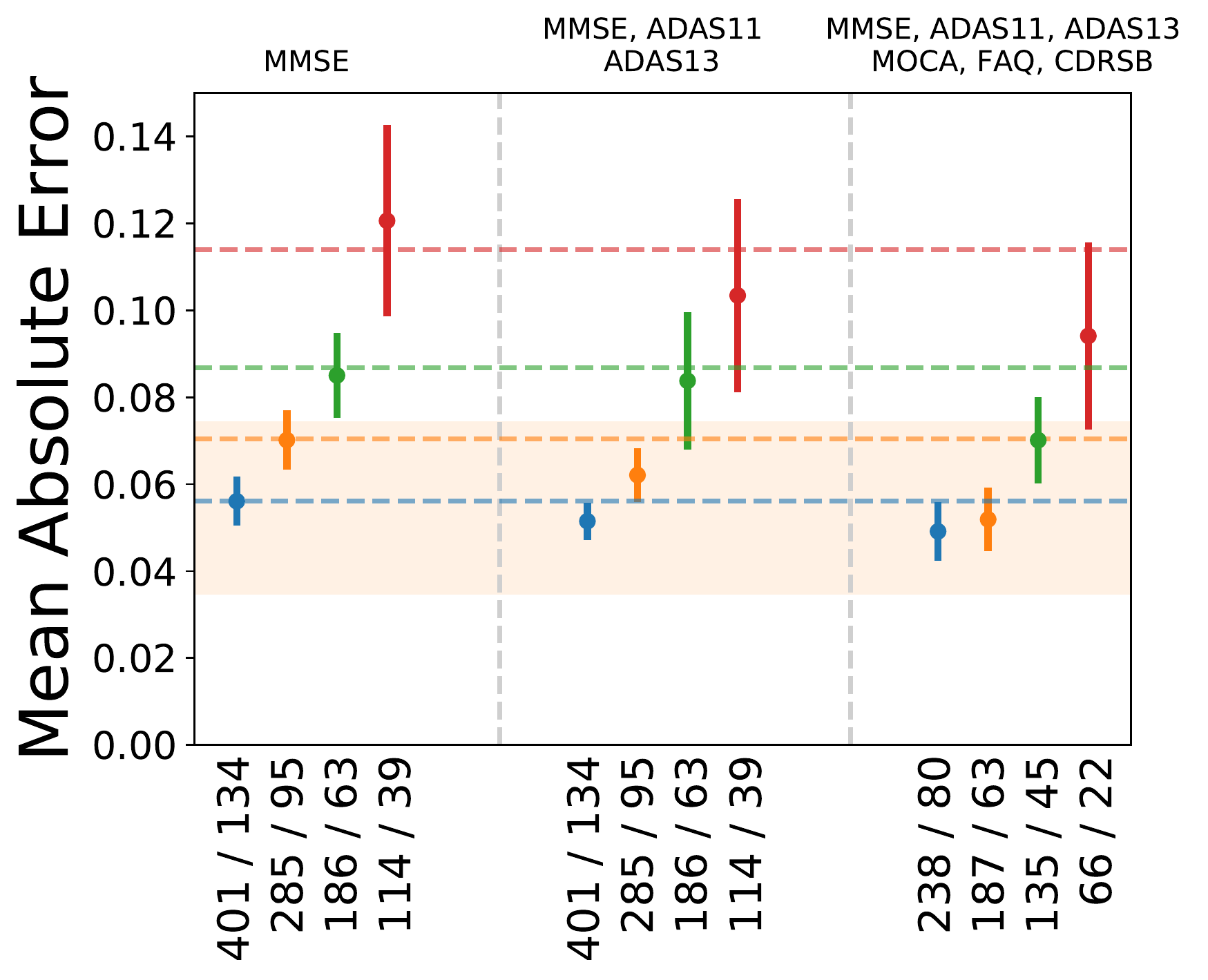}
      \label{fig:standard-setting}
                         }
    \subfloat[Augmented data]{
      \includegraphics[width=0.5\textwidth]{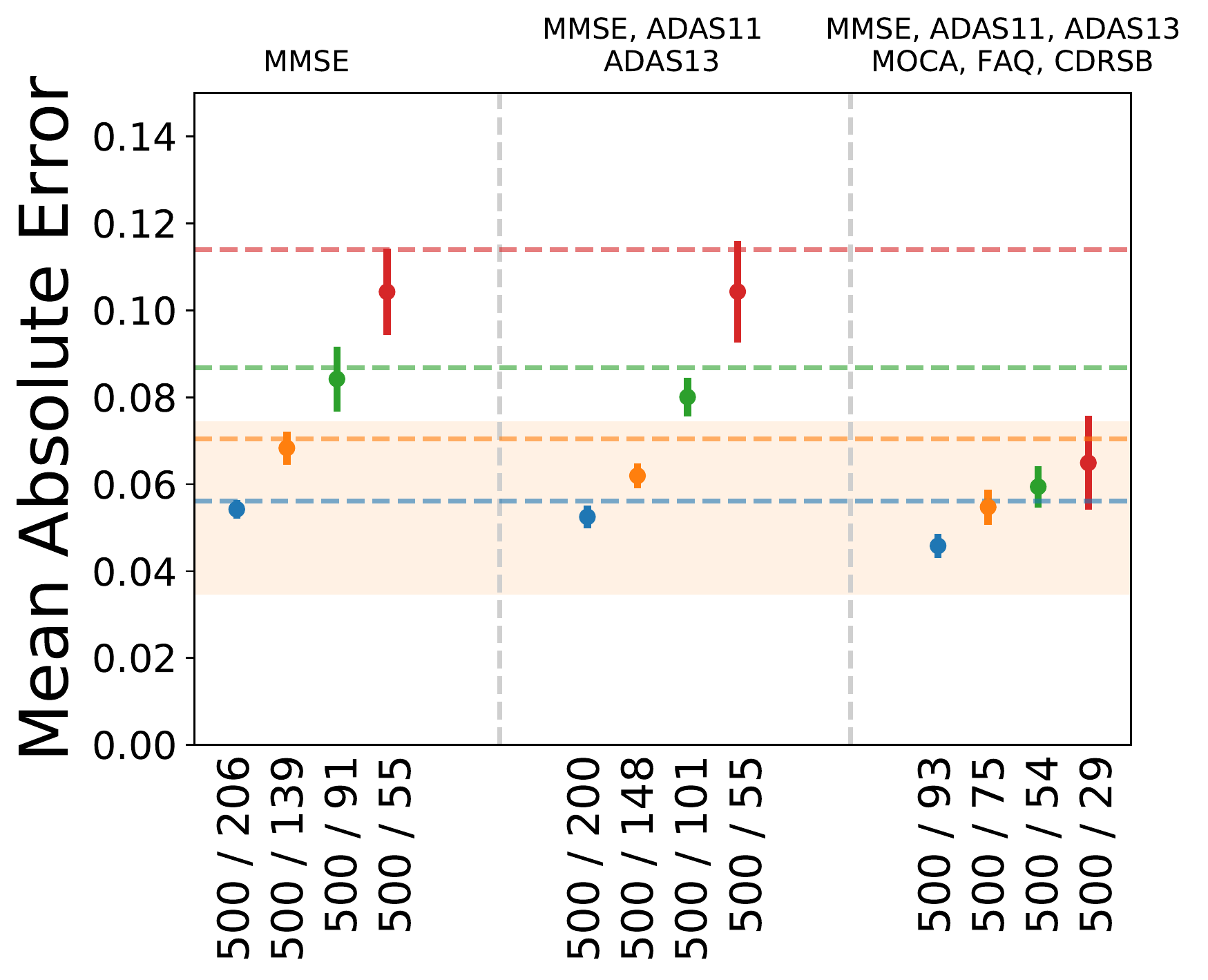}
      \label{fig:augmented-setting}
                         }
    \caption{Prediction of the MMSE in 1 (blue), 2 (orange), 3 (green) and 4 (red) years with different sets of variables (upper part of each column). The colored dashed lines corresponds to the error for the corresponding constant prediction. The pale orange area corresponds to the noise interval in the data. The number at the bottom presents the training and test set sizes. }
    \label{fig:algorithm-comparison}
  \end{center}
\end{figure}

\begin{figure}[h]
  \begin{center}
    \subfloat[MMSE prediction in 3 years]{
      \includegraphics[width=0.5\textwidth]{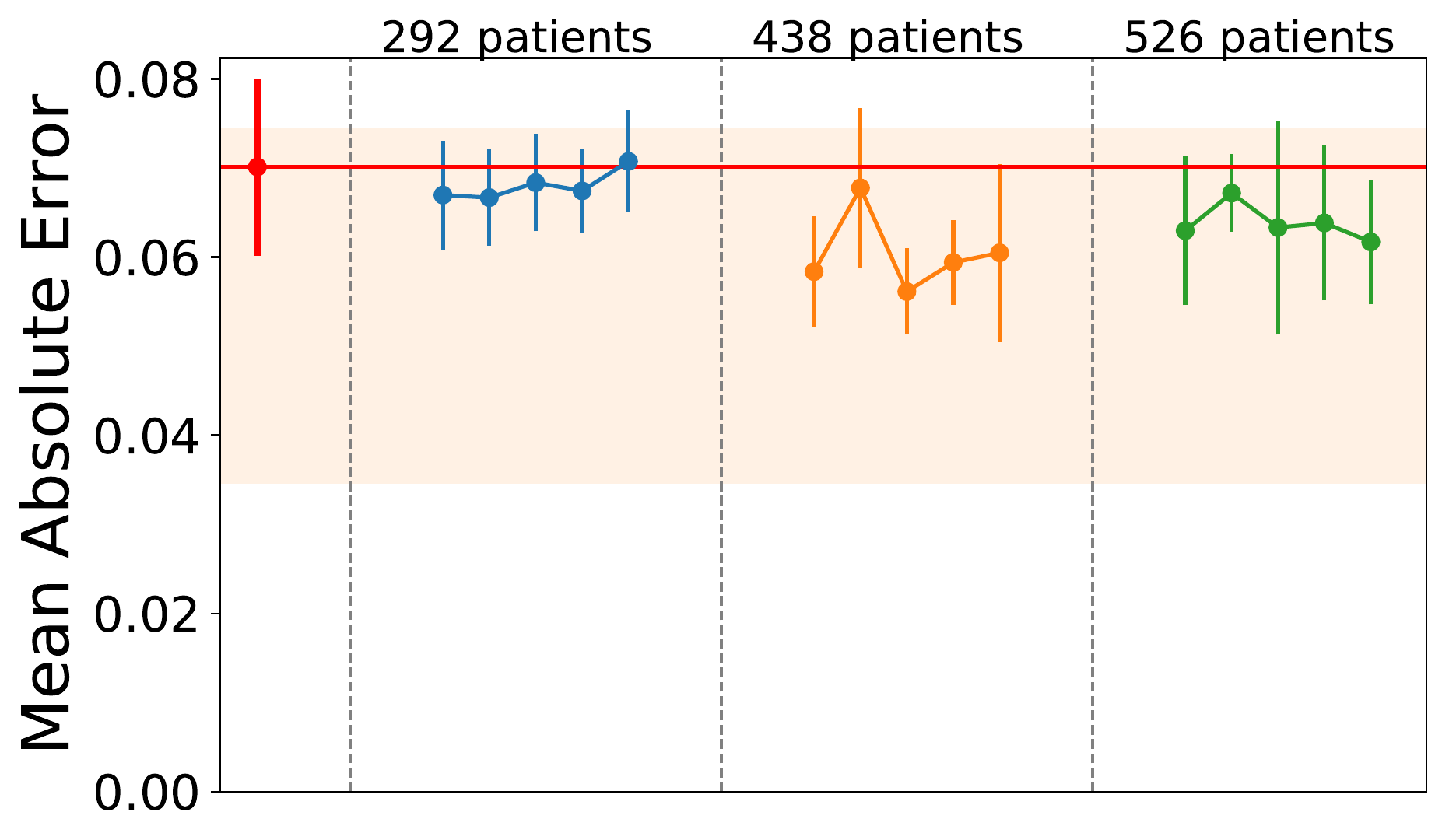}
      \label{fig:three-years-prediction}
                         }
    \subfloat[MMSE prediction in 4 years]{
      \includegraphics[width=0.5\textwidth]{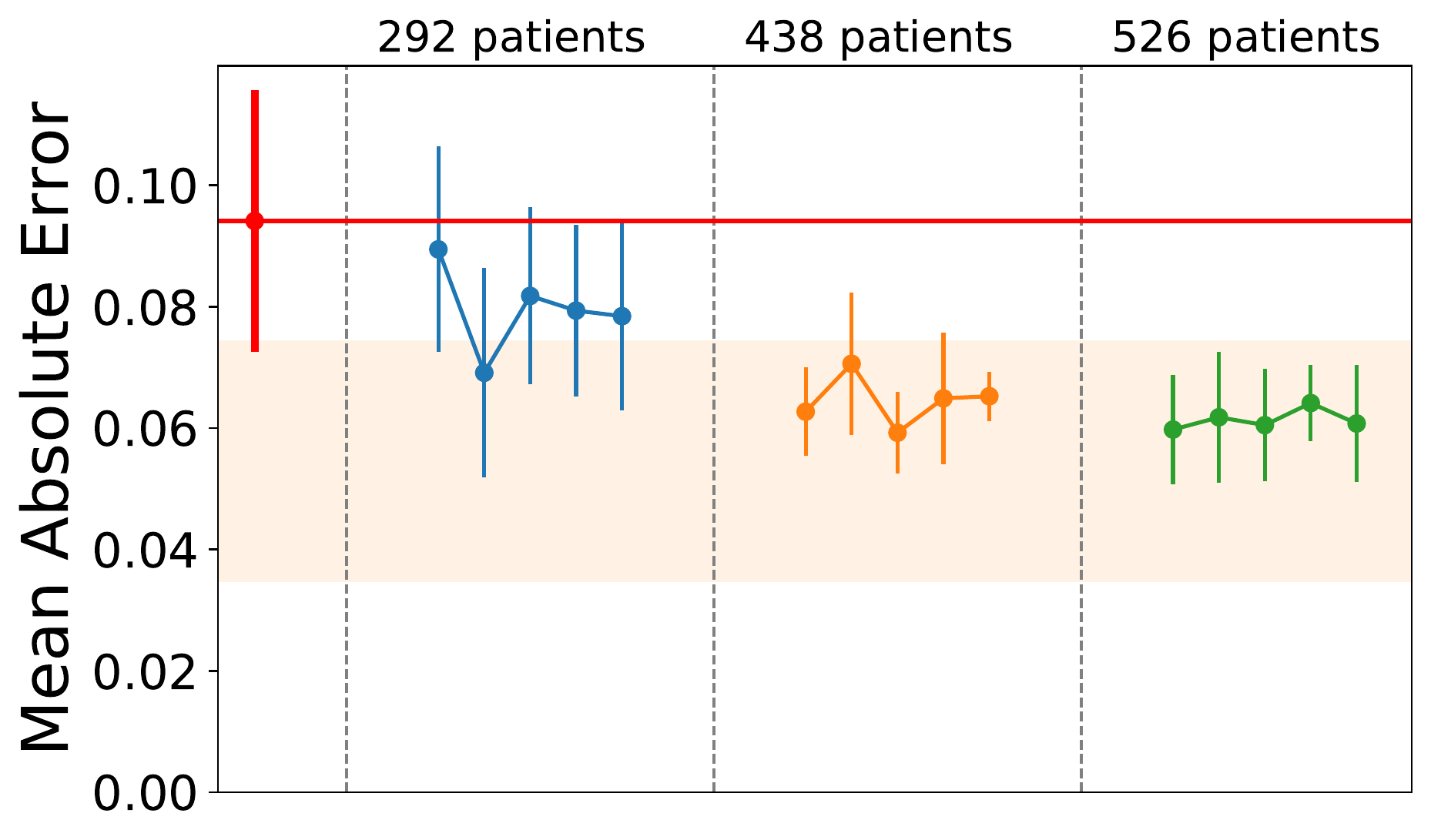}
      \label{fig:four-years-prediction}
                         }
    \caption{MMSE prediction based on MMSE, ADAS-11, ADAS-13, MOCA, FAQ and CRDSB. The red value on the left corresponds to the MAE without simulated data. Then, each column corresponds to a different size of the estimation set. Within each column, we simulate, from left to right, 50, 100, 250, 500 and 1000 virtual patients.}
    \label{fig:prediction-accuracy}
  \end{center}
\end{figure}

The results report the Mean Absolute Error (MAE). To estimate the variance of the estimation procedure, the results are presented with error-bars corresponding to the mean and standard deviation of 10 independent runs with different test splits. The results are compared to, first, the constant prediction, i.e. the hypothesis that there is no change of MMSE within the time interval, and, on the other side, the noise in the data. For the MMSE, \cite{clark1999variability} reports two noise values : a standard deviation of 1.3 and 2.8  (out of 30) for respectively cognitively normal and MCI patients. Once normalized and converted to absolute values, it corresponds to MAE errors of 0.035 and 0.074, represented by a pale orange interval on Figures \ref{fig:algorithm-comparison} and \ref{fig:prediction-accuracy}.

\subsection{Prediction of MMSE}

The prediction accuracy in the standard prediction setting, without simulated patients, are presented on Figure \ref{fig:standard-setting}. The different columns correspond to different sets of markers used as input. It is possible to reach noise level prediction up to 2 years in advance with the MMSE, ADAS-11, ADAS-13, MOCA, FAQ and CDRSB. At 3 and 4 years, the prediction, although often better than the constant prediction, is still larger than noise level.

Replacing the training set by a simulated cohort leads to a significant improvement of the prediction, as first shown on Figure \ref{fig:prediction-accuracy}. It corresponds to a decrease of the MAE of 20\% (resp. 37\%) for prediction 3 years (resp. 4 years) in advance. As the part of the patients used in the estimation set may vary, we tested different scenarios that lead to better results when more patients were used. On the contrary, the number of simulated patients does not seem to have a great impact on the quality of the prediction. A possible but preliminary explanation lies in the fact that even though there are not a lot of simulated patients, they already incorporate more (simulated) visits than real patients.

\subsection{Fair comparison of algorithms}

To better exhibit the problem of comparison between different prediction settings, we refer to Figure \ref{fig:standard-setting} where the lower figures represent the number of training and test patients in the related model. These numbers decrease for longer time to prediction but also for different sets of features, as not all the examinations have been assessed at each patients' visit. 

To solve this issue, we simulated 500 virtual patients for each scenario and estimated the MAE on real patients, as shown on Figure \ref{fig:augmented-setting}. The prediction at 3 and 4 years on the left column corresponds to the values of Figure \ref{fig:prediction-accuracy}. The first result to notice, comparatively to \ref{fig:standard-setting}, is that the MAE variance over the 10 runs is reduced, probably due to the increased test set. More interestingly, the predictive power of the ADAS-11, ADAS-13 and MMSE is not better than with the MMSE alone, a result that could not have been stated from the standard prediction. It essentially means that the MMSE alone is a predictor as good as the three variables but needs more patients to train the model on. In the same spirit, FAQ, MOCA and/or CDRSB provide substantial information that allow to reach noise level prediction up to 4 years in advance.

\section{Discussion}

We proposed a data augmentation technique for small data sets that allow to increase the accuracy of the MMSE prediction for MCI subjects. We believe this technique to be a milestone in the ability to accurately compare various algorithms and features for different time to prediction, as it helps simulating training cohort that are comparable in terms of number of subjects, number of visits per subject and overall follow-up duration.

In this regard, we need to further evaluate the simulation procedure by measuring, for instance, the impact of the number of visits simulated, the time-interval between them, or the selection of the first visit. This could benefit other studies by providing a more accurate comparison of the predictive quality of models or new biomarkers. Overall, it gives a better idea of the generalisation errors of such predictive algorithms in a real clinical setting.

\vspace{5mm}

\textbf{This work has been partly funded by ERC grant N\textsuperscript{o}678304, H2020 EU grant N\textsuperscript{o}666992, and ANR grant ANR-10-IAIHU-06.}

\bibliographystyle{splncs04}
\bibliography{biblio}

\end{document}